\documentstyle[12pt]{article}

\begin{document}

\title{On The Construction of Generalized Coherent States for Generalized Harmonic Oscillators}
\author{}
\date{}
\maketitle

\center{\bf M. EL BAZ}\footnote{E-mail address: moreagl@yahoo.co.uk}

\center{\it Facult\'e des sciences, D\'epartement de Physique, LPT-ICAC, \linebreak Av. Ibn Battouta, B.P. 1014, Agdal, Rabat, Morocco}
\vspace{0.5cm}
\center{\bf Y. HASSOUNI}\footnote{E-mail address: Y-hassou@fsr.ac.ma}

\center{\it Facult\'e des sciences, D\'epartement de Physique, LPT-ICAC, 
\linebreak Av. Ibn Battouta, B.P. 1014, Agdal, Rabat, Morocco
\linebreak and
\linebreak the Abdus Salam International Centre for Theoretical Physics
\linebreak strada costiera 11 , 34100 Trieste, Italy}
\vspace{0.5cm}
\center{\bf F. MADOURI}\footnote{ E-mail address: Fethi.Madouri@ipeit.rnu.tn}

\center{\it Facult\'e des sciences, D\'epartement de Physique, LPT-ICAC, \linebreak Av. Ibn Battouta, B.P. 1014, Agdal, Rabat, Morocco \linebreak and \linebreak I.P.E.I.T, 1008 Monfleury, Tunis, Tunisia}

\vspace{1cm}
\abstract{A dynamical algebra ${\cal A}_q$, englobing many of the deformed harmonic oscillator algebras is introduced. One of its special cases is extensively developed. A general method for constructing coherent states related to any algebra of the type ${\cal A}_q$ is discussed. The construction following this method is carried out for the special case.}

\vspace{0,3cm}

\pagebreak

\section{Introduction}

The construction of coherent states (C.S) is based, following Klauder\cite{klauder}, on three conditions:
\renewcommand{\theenumi}{\alph{enumi}}
\begin{enumerate}
\item Normalisability;
\item Continuity;
\item Resolution of the unity operator;
\end{enumerate}

Many attempts have been made to build out different families of C.S., the general approach followed by most of the authors on this subject is to design special functions generalizing the exponential function. The conventional bosonic C.S are indeed obtained by involving the exponential function which satisfies the equation $ \displaystyle {df \over dz}=f(z)$. These states are, of course, eigenstates of the annihilation operator. In the framework of the deformed lie algebras and groups, the question then, is how can one define C.S corresponding to physical systems having such particular groups of symmetry. Many authors followed the same approach to construct C.S related to deformed harmonic oscillators. The main difficulty which arises is to find a suitable resolution of unity. In fact, the first two conditions; normalisability and continuity, are trivial. The authors were interested thus in discussing, in a not systematical fashion, how to put the equation allowing the obtention of the conventional exponential function in a convenient form. Some approaches are based on some sort of an equation involving, in addition to the complex variable z, a complex parameter $q$; this latter is nothing but the one used to introduce the deformed harmonic oscillator; $\displaystyle {df(q,z)\over dz} = f(q,qz)$\cite{pen1}. In other cases, one acts  on the definition of the derivation operator: $ \displaystyle{d_qf(z)\over dz} = f(z)$ \cite{per96}. 

In this paper we try to give a new expression of the deformed exponential function allowing the construction of the C.S without changing the notion of derivation. The so obtained C.S will correspond to a q-deformed oscillator that could be considered as the most general one. Particular attention is going to be paid to the discussion of the cases when $q$ is generic and also when it is a phase. We will demonstrate that the construction of the C.S is largely defined for $q$ being a phase.

This paper is organized as follows: In the first section we introduce an algebra ${\cal A}_q$, that seems to englobe many of the q-deformed harmonic oscillator algebras. This algebra is a deformation of the algebra treated in \cite{gazeau1}.Particular attention will be paid to a special q-deformed algebra, that appears as a particular case of the algebra ${\cal A}_q$.

The second section, is devoted to the construction of C.S for the algebra ${\cal A}_q$. The major difficulties related to this approach will be discussed. These difficulties will be overcome for the special case and then the construction achieved.

\section{Deformed Harmonic Oscillator}
\subsection{ Preleminary }
This section concerns the description of the harmonic oscillator algebra seen in its general form. It is given through the one introduced in \cite{gazeau1} ${\cal A}$, which is freely generated by the triplet $\{ I, a, a^{+} \}$, and we have the following commutation relations:
\[
[ a , a^+ ] =  \Delta ' 
\]
\begin{equation}
[ a , \Delta ]  =  \Delta '  a 
\end{equation}
\[
[ \Delta , a^+ ] = a^+ \Delta '
\]
\[
 \vdots 
\]
where the dots mean that one can construct other operators (say of the form $\Delta ''$ by considerning the commutator of the operators $a$ and $a^+$ with the $\Delta$'s. $ \Delta = a^+ a $, and $ \Delta'$ an operator that seems to be the derivative of the operator $ \Delta $ and commuting with it. Indeed the term derivative used in this context, can be justified by considering an operator $ f(N)$ (a given function on the number operator) appearing in the literature of the oscillator algebra. To substantiate this idea let us consider the Fock representation of a statistical system, then the action of $\Delta '$ can be seen as $f(N+1) - f(N)$ which can be seen as a formal derivative of the operator $\Delta \equiv f(N)$. We recall that by considering the Fock representation of the algebra (1), many realizations can be obtained, we introduce one of them as follows:

\begin{eqnarray}
a|n> & = & \sqrt {\delta _n} |n-1> \nonumber \\
a^+|n> & = & \sqrt {\delta _{n+1}} |n+1> \nonumber \\
\Delta |n> & = & \delta _n |n> \\ 
\Delta' |n> & = & (\delta _{n+1} - \delta _n ) |n> \nonumber
\end{eqnarray}
The spectrum of $\Delta$ forming a strictly increasing sequence of positive numbers; (1) is the q-oscillator algebra when the sequence is the set of q-deformed natural numbers \cite{gazeau1}; by considering the Fibonnacci sequence the same authors have also obtained a new type of quantum algebras.

Let us now discuss promptly how to obtain the ordinary cases from ${\cal A}$, by means of an adequate choice of the  function $\delta _n$:
\begin{itemize}
\item $ i) $ By putting $ \delta _n = n $, we obtain the usual bosonic harmonic oscillator:
\[ 
a^+a = N \;\;\; \mbox{and} \; \;\; aa^+ = N+I
\]
we have then:
\begin{equation}
[ a , a^+ ] = I = \Delta' = (N+I) - (N)
\end{equation}
and the higher order derivatives of $\Delta$ cancels out.

\item $ii)$ the standard q-deformation of the harmonic oscillator \cite{arik}:
\begin{equation}
aa^+ - qa^+a = I
\end{equation}
is obtained by putting:
\begin{equation}
\delta_n = {{1 - q^n}\over {1-q}}
\end{equation}
\item $iii)$ the case that is of interest for us in this paper is \cite {yas1}:
\begin{equation}
aa^+ - qa^+a = q^{-N}
\end{equation}
is obtained through:
\begin{equation}
\delta _n = { q^n - q^{-n} \over q - q^{-1}}
\end{equation}

\end{itemize}

In what follows we shall present a new and more general form of the dynamical algebra ${\cal A}$ (the generalization here is in the sense of deformation).

\subsection{ Deformed Dynamical Algebra:}
As for ${\cal A}$, we define an algebra ${\cal A}_q$ that is freely generated by the triplet $\{I, a, a^+ \}$, the difference here is that the law from which the algebra is defined, is a "q-mutator" instead of a commutator:

\[
[a,a^+]_q  =  aa^+ - qa^+a = \Delta'_q 
\]
\begin{equation}
[a,\Delta _q]_q  =  a\Delta _q - q\Delta _qa = \Delta'_qa 
\end{equation}
\[
[\Delta _q,a^+]_q =  \Delta_qa^+ - q a^+\Delta _q = a^+\Delta' _q 
\]
\hspace{6cm}$ \vdots  $

where $ \Delta_q = a^+a$ and $q$ a generic complex parameter.

By analogy with $\cal A$,  $\Delta' _q$ is interpreted as a q-derivative of ${\Delta }_q$ and is also commuting with it.

In order to recover the known q-deformed oscillators, well established in the literature, we have to proceed in a  special manner. The goal is actually to overcome the problems arising from the technical difficulties related to the classification of the deformed oscillator algebras. Our purpose led us to consider the action of the algebra (9) on the Fock space:

\begin{eqnarray}
a|n> & = & \sqrt{{[\rho_n]}_q} \;|n-1> \nonumber \\
a^+ |n> & = & \sqrt{{[\rho_{n+1}]}_q} \; |n+1> \nonumber \\
\Delta_q |n> & = & {[\rho_n]}_q \; |n> \\
\Delta'_q |n> & = & ({[\rho_{n+1}]}_q -q{[\rho_n]}_q) \; |n> \nonumber
\end{eqnarray}

$\Delta_q$ and its derivatives are diagonal operators depending on the number operator $N$. 

The original case is ensured by taking q going to 1 and replacing ${[\rho_n]}_q$ by $\delta_n$, one recovers the algebra $\cal A$ in (1). We notice that the cases $ii)$ and $iii)$ can be obtained directly from the algebra ${\cal A}_q$  by taking $\rho_n = n$, and using the convenient "box" function $[ \;\;\;]_q$. What makes ${\cal A}_q$ more interesting than $\cal A$ is that these special cases can be obtained directly from the algebra; i.e. without having to consider its realization in a Fock space, (in contrast to the algebra $\cal A$ where we can't obtain these special cases without considering this realization); and this is very important from a  mathematical point of view.

The physical meaning of these type of algebras is related to the description of intermediate statistics. In this work, we try to introduce a consistent method allowing the construction of any known oscillator algebra by simply getting it from ${\cal A}_q$ in (8). In this way we recover, especially, the algebra ${\cal A}$ in (1) by taking q equal to 1.

\subsection{Q-oscillator algebra:}

We have proved in a previous work \cite{yas1} that the algebra (6) is the most general one within those encountered in our works \cite{yas2, yas3, yas4}. For this reason we try to perform a passage from ${\cal A}_q$ to (6) by means of a change of operators. Firstly we recall that the algebra (6) is generated by $\{A, A^+, N \}$ obeying the Q-mutation relations:
\begin{equation}
{[A, A^+ ]}_Q = AA^+ - QA^+ A = Q^{-N}
\end{equation}

In \cite{yas1}, it has been proved that (6) generalizes many q-deformed oscillator algebras and also shown that it gives interesting properties for $Q$ going to a root of unity. In particular, quantum algebras and superalgebras were constructed out from it.

Now we investigate the way leading to the algebraic relations (10) from (8). This can be done by involving an operator F in the change of operators as follows:
\begin{eqnarray}
A &=& Fa \nonumber \\
A^+ &=& \alpha a^+ F 
\end{eqnarray}

$\alpha$ being an arbitrary complex number. Notice that in the case of deformation the operation $"^+"$ is not necessary the complex conjugation of generators but just becomes so for $Q=1$.
We impose, in order to be able to obtain the algebra (10), on $F$ to be diagonal and to satisfy $ F=F^+ $.

On the Fock basis, $F$ acts as follows:
\begin{equation}
F|n> = f_n |n>
\end{equation}
It is easy to check that $f_n$ is subject to the following constraint:
\begin{equation}
\sqrt{{[n+1]}_Q} = f_n \; \sqrt{{[{{\rho}_{n+1}}]}_q}
\end{equation}
where we have used the action of the generators in (10) on a Fock space basis:
\begin{eqnarray}
A|n> &=& \sqrt{{[n]}_Q} |n-1> \nonumber \\
A^+ |n> &=& \sqrt{{[n+1]}_Q}|n+1> \\
A^+A = {[N]}_Q &=& {Q^N -Q^{-N} \over Q-Q^{-1}} \nonumber
\end{eqnarray}
In terms of operators, $F$ can be written as:
\begin{equation}
F^2 = {[N+1]}_Q {[\rho _{N+1}]}^{-1}_q
\end{equation}
but we require that the operator $\rho _N$ must be invertible.

To conclude this paragraph, we notice that the aim of this study is to show that it is possible to classify the deformations of the oscillator algebras if we define correctly the statistics we are interested in. On the pure mathematical background, work is under progres and we will show in a forthcoming paper that the inherent mathematical properties are well established. It will be shown that it is indeed possible to get a consistent unification of the deformed oscillator algebras. It will become clear that given a consistent statistics we will find systematically the corresponding C.S.

\section{Coherent States:}
\subsection{Definition:}

In this section we are going to construct C.S for the harmonic oscillator (10), by deducing them from the ones related to the algebra ${\cal A}_q$ in (8). The states constructed are "coherent" in the sense of Klauder \cite{klauder}. In this definition C.S are vectors in some Hilbert space satisfying the following properties:

\begin{enumerate}
\item Normalisability:
\begin{equation}
<z|z> = 1
\end{equation}
\item Continuity in the label z:
\begin{equation}
\mbox{if }\;\;\;\;\;|z-z'|^2  \to 0 \;\;\;\;\; \mbox{then }\;\;\;\;\; ||z>-|z'>|^2 \to 0
\end{equation}
\item Resolution of unity:

There must exist a positive weight function $W$ such that:
\begin{eqnarray}
\int \int d^2z |z> W(|z|^2)<z| &=& I \nonumber \\
&=& \sum_{n\ge 0} |n><n|
\end{eqnarray}
where $ z = \alpha +i\beta $ and $ d^2z = d\alpha d\beta$
\end{enumerate}

\subsection{Construction:}
We begin by trying to construct C.S for the algebra ${\cal A}_q$. A trivial way to get eigenstates of the annihilation operator $a$, that are going to be good candidates as coherent states in the sense mentioned above is:
\begin{equation}
||q,z> = \sum ^{\infty}_{n\ge 0} {z^n \over \sqrt{{[\rho _n]}_q!}}|n>
\end{equation}
where the factorial function is defined by:
\begin{eqnarray}
{[\rho _n]}_q! &=& {[\rho _{n}]}_q {[\rho _{n-1}]}_q \dots {[\rho _1]}_q \\
&\mbox{and}& \nonumber \\
{[\rho _0]}_q! &=& 1 \nonumber
\end{eqnarray}

A straightforward calculation shows that indeed (19) are eigenstates of the annihilation operator:
\begin{equation}
a||q,z> = z||q,z>
\end{equation}
we rewrite these states in the following useful forms:

\begin{eqnarray}
||q,z> &=& \sum _{n\ge 0}{{z^n \over {[\rho _n]}_q!}{(a^+)}^n} |0> \nonumber \\
&=& exp _q (za^+) |0> 
\end{eqnarray}

where we have introduced the deformed exponential function through:
\begin{equation}
exp _q (x) = \sum _{n \ge 0}{x^n \over {[\rho _n]}_q!}
\end{equation}

At this level it will be convenient to make few restrictions on some of the entities we are dealing with. In fact, we shall impose the following condition on the function${[{\rho _n}]}_q$:
\[
{[{\rho _n}]}_q =\bar {[{\rho _n}]}_q
\]
where the bar means complex conjugation.

This is indeed the case for the case we are considering in this paper i.e (10) for $|q| =1$ .

This choice will be motivated in what follows.

Now, let's check if these states do satisfy relations a), b) and c), and hence are C.S \cite{kps}.

a)
The states $||q,z>$ (as any vector in a Hilbert space) are normalisable. Indeed we have:
\begin{eqnarray}
<q,z||q,z> &=& \sum _{n\ge 0}{{|z|}^{2n} \over {[\rho _n]}_q!} \nonumber \\
&=& exp _q (|z|^2)
\end{eqnarray}
The normalized states are then given by:

\begin{eqnarray}
|q,z> &=& {\cal N}({|z|}^2) ||q,z> \nonumber \\
&=& exp _q (- {{|z|}^2 \over 2}) ||q,z> \\
&=& exp _q (- {{|z|}^2 \over 2}) exp _q (za^+) |0> \nonumber
\end{eqnarray}

The overlap term of two such states is given by:
\begin{eqnarray}
<q,z|q,z'> &=& {\cal N}({|z|}^2){\cal N}({|z'|}^2) \sum _{n\ge 0} {{\bar z}^nz^n \over {[\rho _n]}_q!} \nonumber \\
&=& exp _q(-{{|z|}^2 \over 2}) exp _q (-{{|z'|}^2 \over 2}) exp _q ({\bar z}z')
\end{eqnarray}

As is clear from what precedes, we wouldn't be able to write the formulas in (24,25,26) in such an aesthetic form, without the condition imposed on ${[{\rho _n}]}_q$. However, we can obtain analoguos formulas for a generic ${[{\rho _n}]}_q$, but we will lose this resemblance to the conventional case; i.e the bosonic C.S.

In particular for (24) we would get:
\[
<q,z||q,z> = \sum _{ n\ge 0} {|z|^{2n} \over \sqrt{{[\rho_n]}_q!} \sqrt{{[\rho_n]}_{\bar q}!}}
\]

so we couldn't involve the deformed exponential function!

b)

The continuity of the states $|q,z>$ is deduced from the continuity of the overlap term (26) by remarking that:
\begin{equation}
\big | |q,z> - |q,z> \big |^2 = 2(1-Re<q,z|q,z'>)
\end{equation}

c)
What is relatively complicated is the third condition; the resolution of unity. In fact we must find the weight function $W(|z|^2)$ such that (18) holds:

using (25) in (18) we obtain:
\begin{equation}
\int \int d^2z \sum _{n,m\ge 0}{z^n\over \sqrt {{[\rho _n]}_q!}} {{\bar z}^m \over \sqrt{{[\rho _m]}_q!}}{\cal N}^2({|z|}^2)|n><m| W({|z|}^2) = I
\end{equation}
using $ d^2z = d\alpha d\beta = rdrd\theta \;\;\;,$ where $\;\;\; z=\alpha + i\beta = r e^{i\theta}$, (28) is rewritten as:
\begin{equation}
\sum _{n,m\ge 0} \bigg\{ { 1 \over \sqrt{{[\rho _n]}_q!}\sqrt{{[\rho _m]}_q!}} \int _0^{\infty} rdrr^{n+m}{\cal N}^2 (r^2)W(r^2) \int _0^{2\pi} d\theta \;e^{i\theta (n-m)}|n><m|\bigg \} = I
\end{equation}
which implies:
\begin{equation}
\sum _{n\ge 0} \bigg \{ {2\pi \over {[\rho _n]}_q!} \int rdr r^{2n}{\cal N}^2(r^2)W(r^2)|n><n| \bigg \} = I
\end{equation}
using the change of variable $ x= r^2 $ it becomes:
\begin{equation}
\sum _{n\ge 0} { \pi \over {[\rho _n]}_q!} \bigg \{ \int dx\; x^n \; {\cal N}^2(x) W(x) \bigg \} |n><n| = I
\end{equation}
this equation, together with the unity equality $\displaystyle \sum _{n\ge 0} |n><n|$ implies the following condition on $W(x)$:
\begin{equation}
\int dx \; x^n \; {\cal N}^2(x) W(x) = {{[\rho _n]}_q! \over \pi }
\end{equation}
or also;
\begin{equation}
\int dx \;x^n \; \tilde{W}(x) = {{[\rho _n]}_q! \over \pi }
\end{equation}
here we put:
\begin{equation}
\tilde{W}(x) = {\cal N}^2(x) W(x)
\end{equation}
equation (33) is the well known power-moment problem \cite{akh}. As stated in \cite{kps}, this equation don't have solutions for a general function in the r.h.s. here $ {[\rho _n]}_q$. Indeed for such solutions to exist this function have to satisfy some conditions \cite{kps,akh}. However in practice it is hard to prove that a given function do satisfy such conditions\footnote{ we would have encountered the same problem if we tried to construct C.S related to $\cal A$ since the construction would have been the same, all we would have to do is to change ${[\rho _n]}_q$ by $\delta _n$, but even then we can't tell if (33) have solutions or not, sine $\delta _n$ still general!.}.

To overcome this difficulties, some physicists get the ingenious idea of attacking the problem from another angle. Namely they look for adequate, known solutions for the moment problem (33); i.e for ${[\rho _n]}_q$'s for which (33) have known solutions. Then using the same reasoning; but in the opposite way this time; they construct the deformed algebra (deformed bosons) for which the C.S 's resolution of unity would have led to (33), with the correspending ${[\rho _n]}_q $. For more details (nothing like the original!)\cite{kps}.

It is worth to mention that many important papers adopting this point of view appeared during the last 2 years \cite{pen1,kps,pen2,pen3}. However we are not going to use this approach, simply because it won't provide us with the desired C.S.

As mentioned above no solutions exists for a genric ${[\rho _n]}_q$, therefore no C.S exists; at least in the form (25) and in the sense of Klauder; for the corresponding deformed bosons. However for ${[\rho _n]}_q $'s  for which (33) have solutions the states (25) are coherent!

In particular for ${[\rho _n]}_q = n$(i.e ordinary bosons) the solution of (33) is: 
\begin{equation}
\tilde W(x) = {e^{-x} \over \pi}
\end{equation}
with $\displaystyle {\cal N} =e^{-x\over 2}$ we get  $\displaystyle W(x) = {1\over \pi}$.

and the resolution of unity is:
\begin{equation}
{1\over \pi} \int d^2z \; |z><z| = \sum _{n\ge 0}|n><n| = 1
\end{equation}
where
\begin{eqnarray}
|z> &=& exp (-{|z|^2 \over 2}) exp(za^+) |0> \nonumber \\
&=& exp(- {|z|^2 \over 2}) \sum _{n\ge 0}^{\infty} {z^n \over \sqrt{n!}} |n>
\end{eqnarray}
are the conventional bosonic C.S.

For $ \displaystyle {[\rho _n]}_q = {1-q^n \over 1-q}$; with $q$ a real parameter; the corresponding solution of (33) and the corresponding C.S were constructed in \cite{quon}:

For the case we are considering in this paper:
\[
{[\rho _n]}_q = {q^n - q^{-n} \over q-q^{-1}}
\]

It is evident that this function do satisfy condition ${[{\rho _n}]}_q =\bar {[{\rho _n}]}_q$ when q is a phase. 

We shall find the corresponding function $\displaystyle \tilde W(x)$ satisfying (33). For this, we make use of the Fourrier transforms. We proceed as follows:

We multiply both sides of (33) by $\displaystyle \big({(iy)^n\over n!} \big )$, then summing over $n$ yields:
\begin{equation}
\int _0 ^{\infty} {dx\;e^{iyx}\; \tilde W(x)} = \sum _{n=0}^{\infty}{ {[\rho_n]}_q!(iy)^n \over \pi n!} = \bar W(y)
\end{equation}
To proceed further it is clear that the series in the r.h.s has to converge. This is the case when the deformation parameter q is a phase. Thus in this case we can get the inverse Fourrier transform of $\bar W(y)$:
\begin{equation}
\tilde W(x) = { 1 \over 2\pi} \int _{-\infty}^{\infty}{e^{-iyx} \bar W(y) dy}
\end{equation}
which yields for $W(x)$
\begin{equation}
W(x) = { {\cal N}^{-2}(x) \over 2\pi} \int _{-\infty}^{\infty} {e^{-iyx} \bar W(y) dy}
\end{equation}

and the resolution of unity then is:
\begin{equation}
\int \int d^2z |q,z> W(|z|^2)<q,z| = I 
\end{equation}
where $|q,z>$ are given in (21) and $W$ is given in (40).

We have thus demonstrated that the states (21) are actually C.S. of the deformed harmonic oscillator given in (10) when the parameter of deformation is a phase.

The next step is rather trivial. In fact, using the resolution of unity (41), we can write any state $|\psi >$ in the Hilbert space, in terms of the C.S. (21)\footnote{of course for ${[\rho _n]}_q = {q^n - q^{-n} \over q-q^{-1}}$ and q a phase}:
\begin{equation}
|\psi > = \int d^2z\; W(|z|^2)\;{\cal N}(|z|^2)\; \psi (\bar z) |q,z>
\end{equation}
where
\begin{equation}
\psi(z) = \sum _{n\ge 0} {z^n \over \sqrt{[\rho _n]_q!}}<n|\psi >
\end{equation}
This establishs a one to one correspendence between vectors $\{|\psi>\}$ of the Hilbert space and the analytical functions $\psi(z)$ on the complex plane. This also states the overcompletness of the C.S. (21) \cite{per86}. In fact when $|\psi >$ is a C.S. $|q, z'>$, (42) can be written as:
\begin{equation}
|q,z'> = \int d^2z\; W(|z|^2)\; {\cal N}(|z|^2)\; {\cal N}(|z'|^2)\; exp_q({\bar z}z')|q,z>
\end{equation}
which shows the linear dependence between the C.S (25), and hence their overcompletness.

Moreover we have:
\begin{equation}
||\psi||^2 = <\psi|\psi> = \int {d^2z \; W(|z|^2) \; exp_q (-|z|^2) \;|\psi ({\bar z})|^2}
\end{equation}
and 
\begin{equation}
<\psi _1|\psi _2> = \int d^2z \; W(|z|^2) \; exp_q (-|z|^2) \;{\bar \psi}_1({\bar z}) \psi _2 ({\bar z})
\end{equation}
and
\begin{equation}
\psi ({\bar z}) = \int {d^2z' \; W(|z'|^2)\; exp_q(-|z'|^2)\; exp_q({\bar z}z')\; \psi ({\bar z}')}
\end{equation}

We have thus defined a Bargmann-Fock space (space of entire analytical functions on the complex plane) for our q-oscillator (10), in analogy to the Bargmann-Fock space for the conventionnal harmonic oscillator:

In this space the scalar product is given by (46), in which we recongnize the measure in this space as being $ \displaystyle d^2z W(|z|^2)exp_q(-|z|^2) $ instead of $\displaystyle {d^2z \over \pi} exp(-|z|^2) $ for the bosonic harmonic oscillator. Moreover, in this space, the role of the delta function $\delta (z,z')$ is played by $ \displaystyle exp_q({\bar z}z') $ as clearly seen from (47)(in the conventionnal Bargmann-Fock space this role is played by $ exp ({\bar z}z')$.

\section{Addendum}

In this paragraph we give some explanations and remarks on some of the
entities introduced and used in this paper  without which the use of these entities will be a non-sens.

\begin{itemize}
\item The deformed exponential function introduced in (23) is strongly
convergent for the case we are considering (6,7), and its convergence radius
is $\infty $. This reflects its self in the fact that the C.S constructed are
defined on the whole complex plane, this in contrast to \cite{arik} where the
C.S were defined on a disk: $ |z| < $ some $R_q$.

\item the imposed condition $ [\rho _n]_q = \bar {[\rho _n]_q }$, is
essentialy for aesthetic reasons, and the construction could be carried out
without this requirement.

\item The term $exp_q(-t)$ appearing in this paper (such as in eq(25,26)
or implicitely in eq(27)), in no way means deformed exponential function  of
$(-t)$. This is a (may be missleading) notation we have used to keep the
analogy with the classical case, and it stays for $ 1/exp_q(t)$, which for the
case we consider here is different from $exp_q(-t)$.

\item the same remark as above stands for $exp_q(t/2)$, this means
$ \bigg( exp_q(t)\bigg )^{1\over 2}$.

\end{itemize}

the last two remarks are due to a drastical difference between the deformed
function used here and the usual exponentiel function, namely:
\[
exp_q(x)exp_q(y) \neq exp_q(x+y)
\]

\pagebreak
\section{Conclusion}
The algebra ${\cal A}_q$ discussed in this paper generalizes, to some extent, many deformed harmonic oscillator algebras. Thus one can get C.S asociated to these deformed algebras, as the corresponding particular cases of the C.S associated to the algebra ${\cal A}_q$. Still that there's no method for constructing such C.S. A tentative was made in this paper but due to technical problems, such a construction couldn't be carried out to the end; we had to specify which particular algebra of ${\cal A}_q$ we are interested in; when such a choice was made the construction was possible.

Many perspectives opens up for this work, one should try to study the deformed algebra ${\cal A}_q$, introduced in this paper, its algebraic structure, its representations... One can also try to use the same approach used in this wok for the construction of C.S. for other particular algebras of ${\cal A}_q$.

Concerning the particular algebra (10) discussed in this paper it will be interesting to analyse the drastical difference between the two cases $|Q|=1$ and a generic $Q$. It will be also interesting to try to extract complet subsystems of the C.S (21) in the spirit of what was done in \cite{per96}.

\section{ Aknowledgment }
The authors wish to thank the Abdus Salam International Centre for Theoretical Physics, Trieste, Italy, where part of the work was achieved.

\end{document}